\documentclass[aps,prl,twocolumn,floatfix,groupedaddress,showpacs]{revtex4-1}

\usepackage{amsmath}
\usepackage{graphicx}
\graphicspath{{/}}
\usepackage[breaklinks=true,colorlinks=true,allcolors=blue]{hyperref}

\begin{document}


\title{Self-Bound Quantum Droplet with Internal Stripe Structure in 1D Spin-Orbit-Coupled Bose Gas}


\author{Yuncheng Xiong}
\author{Lan Yin}
\email[]{yinlan@pku.edu.cn}
\affiliation{Peking University}

\date{\today}

\begin{abstract}
We study the quantum-droplet state in a 3-dimensional (3D) Bose gas in the presence of 1D spin-orbit-coupling and Raman coupling, especially the stripe phase with density modulation, by numerically computing the ground state energy including the mean-field energy and Lee-Huang-Yang correction.  In this droplet state, the stripe can exist in a wider range of Raman coupling, compared with the BEC-gas state.  More intriguingly, both spin-orbit-coupling and Raman coupling strengths can be used to tune the droplet density.
\end{abstract}


\maketitle

\paragraph{Introduction.} Ultracold atoms have been excellent platforms for investigating many-body quantum phenomena since the experimental realization of Bose-Einstein condensation(BEC) \cite{anderson1995,bradley1995,davis1995}. In most cases, Gross-Pitaevskii (GP) equations, derived by minimizing mean-field energy functional with respect to condensation wavefunction, provide a good description for the BEC state of trapped Bose gases \cite{dalfovo1999}.  The next-order correction to the gound state energy, i.e. the Lee-Huang-Yang(LHY) energy, is usually negligible in the dilute limit.  However in a binary boson mixture, it was found that when the attractive inter-species coupling constant $g_{\uparrow\downarrow}$ is a little larger in magnitude than the geometric average of the repulsive intra-species coupling constants $g_{\uparrow\uparrow}$ and $g_{\downarrow\downarrow}$,  the repulsive LHY energy overtakes the attractive MF ground-state energy, and the system becomes a self-bound quantum droplet \cite{petrov2015}.
The quantum droplet was first observed in dipolar Bose gases \cite{kadau2016,ferrier2016,chomaz2016,schmitt2016,trautmann2018} and later in binary boson mixtures \cite{cabrera2018,cheiney2018,semeghini2018}. The self-binding mechanism of a single-component dipolar Bose gas is similar to that of a boson mixture except that the residual MF attraction arises from the counterbalance between attractive dipole-dipole interaction and repulsive contact interaction. Theoretically, the quantum droplet has been investigated with various methods, including variational HNC-EL method \cite{staudinger2018}, ab initial diffusion Monte-Carlo \cite{cikojevic2018}, and extended GPEs with the LHY correction included \cite{wachtler2016}.

\par On the other hand, Raman-induced spin-orbit-coupling(SOC) has been realized experimentally in recent years both in bosonic \cite{lin2011,zhang2012} and fermionic \cite{cheuk2012,wang2012} systems. Alternative scheme of SOC which is immune from heating problem has been theoretically \cite{zhang2013,xu2013,anderson2013} and experimentally \cite{luo2016} investigated.  In a two-component Bose gas with a one-dimensional (1D) SOC, a stripe structure appears when the inter-species coupling constant is smaller than geometric average of intra-species coupling constants below a critical Raman coupling (RC) \cite{wang2010,ho2011,li2012,li2013,lin2011,ji2014,li2017}.  In previous studies, the stripe state has been investigated in the BEC-gas region with repulsive MF ground-state energy.  Recently, theoretically studies \cite{sachdeva2020,sanchez2020} reveal that the stripe state can also exist in the quantum droplet regime.  In Ref. \cite{sachdeva2020}, the quantum droplet was found in a two-dimensional Bose gas with very weak SOC, where the LHY energy density was approximated by that of a uniform system without SOC.  In Ref. \cite{sanchez2020} the LHY energy of a three-dimensional system with SOC was calculated numerically, and its fitted form was used in the extended (GP) equation.  The phase transition between a stripe gas and a stripe liquid was found by tuning coupling constants and RC.  In their calculation, the ultraviolet divergence in the expression of LHY energy was removed by dimensional regularization.  In this work, we apply the standard regularization scheme to treat the ultraviolet divergence in the LHY energy of a three-dimensional system with SOC.  We found that the droplet density can be easily tuned by RC and SOC, even to the zero limit.  Compared to the case with a repulsive inter-species interaction, in the quantum droplet the stripe phase can exist in a bigger regime of RC and SOC.
\paragraph{Boson mixture with SOC.} We study a two-component Bose gas system with total particle number $N$ and volume $V$.  In momentum space, its Hamiltonian is given by
\begin{align}
H&=\sum_{\mathbf{k}}\sum_{\rho\rho^{\prime}}\hat{\phi}^{\dagger}_{\rho\mathbf{k}}\bigg(\frac{(\mathbf{k}-k_r\mathbf{e}_x\sigma_z)^2}{2}+\frac{\Omega}{2}\sigma_x\bigg)_{\rho\rho^{\prime}}\hat{\phi}_{\rho^{\prime}\mathbf{k}}
\nonumber\\&+\frac{1}{2V}\sum_{\mathbf{k}_1,\mathbf{k}_2,\mathbf{q}}\;\sum_{\rho\rho^{\prime}}g_{\rho\rho^{\prime}}\hat{\phi}^{\dagger}_{\rho\mathbf{k}_1+\mathbf{q}}\hat{\phi}^{\dagger}_{\rho^\prime\mathbf{k}_2-\mathbf{q}}\hat{\phi}_{\rho^{\prime}\mathbf{k}_2}\hat{\phi}_{\rho\mathbf{k}_1},
\label{eq_H}
\end{align}
where  $\hat{\phi}_{\rho\mathbf{k}}$ and $\hat{\phi}_{\rho\mathbf{k}}^{\dagger}$ are the annihilation and creation operators of the $\rho$-component boson with the momentum $\mathbf{k}$, $\{\rho,\rho^{\prime}\}=\{\uparrow,\downarrow\}$, $k_r$ and $\Omega$ are the strengths of SOC and RC respectively, $\mathbf{e}_x$ is the unit vector in x-direction which is the SOC direction.  For convenience, we set $\hbar$ and the boson mass to be one.  In this paper, we focus on the $\Omega<4E_r$ regime where the lower excitation spectrum of the single-particle Hamiltonian has two degenerate minima \cite{lin2011,li2013}.  For simplicity, the interactions are chosen to be symmetric $g_{\uparrow\uparrow}=g_{\downarrow\downarrow}=g$.  The droplet regime is set by the condition $g_{\uparrow\downarrow}\lesssim -g$ \cite{petrov2015}.

\par To implement Bogoliubov approximation to obtain excitation spectra, and thereby LHY correction, we need to know ground state(GS) wavefunction. To this end, we determine GS by variationally minimizing MF energy. We choose GS ansatz to be superposition of plane waves \cite{li2012,li2013},
\begin{align}
\phi(\mathbf{r})=
\begin{pmatrix}
\phi_{\uparrow}\\\phi_{\downarrow}
\end{pmatrix}
=\sqrt{\frac{N_0}{V}}\sum\limits_m
\begin{pmatrix}
\phi_{\uparrow m}\\-\phi_{\downarrow m}
\end{pmatrix}e^{im\mathbf{k}_1 \cdot \mathbf{r}},
\label{eq_phi}
\end{align}
where $N_0$ is the particle number of the condensate, $\mathbf{k}_1\!\equiv\!( \gamma k_r,0,0)$, $\gamma$ is a variational parameter to be determined. In the lowest order, only $m=\pm 1$ components corresponding to the two minima of the lower single-particle spectrum are relevent \cite{li2012}. However, the periodic stripes induced by the condensation of $\pm\mathbf{k}_1$ will lead to the couplings between the momenta differing from each other by reciprocal lattice vectors. Therefore, it is necessary to include all the components with momenta $\mathbf{K}\pm\mathbf{k}_1$ in the higher order approximations \cite{li2013}, where $\mathbf{K}=2s\mathbf{k}_1$ with $s=0,\pm1,\pm2,\dots$, are reciprocal lattice vectors. In short, the summation is over all the odd integer $m=2s\pm1$ in the region $-C_1\leq m \leq C_1$ where the cutoff $C_1$ is a positive odd number.  The normalization relation is given by $\sum_{\rho,m}|\phi_{\rho m}|^2=1$.

\par In the BEC state, following the Bogoliubov prescription, we replace $\hat{\phi}^{(\dagger)}_{\rho m\mathbf{k}_1}$ by $\sqrt{N_0}\phi^{(\ast)}_{\rho m}+\hat{\phi}^{(\dagger)}_{\rho m\mathbf{k}_1}$ and keep terms up to the quadratic order.  The first-order terms vanish due to the minimization of MF energy.  The number of atoms in the condensation is given by
$N_0=N-\sum\nolimits^{\prime\prime}_{\rho,\mathbf{k}}\hat{\phi}_{\rho\mathbf{k}}^{\dagger}\hat{\phi}_{\rho\mathbf{k}}$
which can be used to rewrite $N_0$ in terms of the total atom number $N$.  The MF energy per particle $\varepsilon_{\scriptscriptstyle M\!F}$ and the Bogoliubov Hamiltonian $H_{\scriptscriptstyle B}$ are given by
\begin{align}
\varepsilon_{\scriptscriptstyle M\!F}=&\sum_{m}\sum_{\rho\rho^{\prime}}\bigg(\frac{k_r^2}{2}(m\gamma-\sigma_z)^2-\frac{\Omega}{2}\sigma_x\bigg)_{\rho\rho^{\prime}}\phi^{\ast}_{\rho m}\phi_{\rho^{\prime} m}
\nonumber\\&+\sum_{m+l=i+j}\;\sum_{\rho\rho^{\prime}}\frac{g_{\rho\rho^{\prime}}n}{2}\phi^{\ast}_{\rho m}\phi^{\ast}_{\rho^\prime l}\phi_{\rho^{\prime} i}\phi_{\rho j},
\label{eq_E_MF}
\end{align}

\begin{widetext}
	\begin{align}
	H_{\scriptscriptstyle B}=&E_{\scriptscriptstyle M\!F}+
	 \sum_{\rho\rho^{\prime}}\sum_{\mathbf{k}}\bigg(\frac{(\mathbf{k}-k_r\mathbf{e}_x\sigma_z)^2}{2}+\frac{\Omega}{2}\sigma_x-\mu\hat{I}\bigg)_{\rho\rho^{\prime}}\hat{\phi}_{\rho\mathbf{k}}^{\dagger}\hat{\phi}_{\rho^{\prime}\mathbf{k}}  \nonumber\\
	&+\sum_{\rho,\mathbf{q}}\sum_{m+l=\alpha+\beta}\frac{g_{\rho\rho}l}{2}\bigg[2\phi^{\ast}_{\rho m} \phi^{\ast}_{\rho l} \hat{\phi}_{\rho\alpha\mathbf{k}_1-\mathbf{q}} \hat{\phi}_{\rho\beta\mathbf{k}_1+\mathbf{q}}
	+2\phi^{\ast}_{\rho m} \phi_{\rho -l}
	(\hat{\phi}^{\dagger}_{\rho -\alpha\mathbf{k}_1+\mathbf{q}}  \hat{\phi}_{\rho\beta\mathbf{k}_1+\mathbf{q}} +\hat{\phi}^{\dagger}_{\rho -\alpha\mathbf{k}_1-\mathbf{q}}  \hat{\phi}_{\rho\beta\mathbf{k}_1-\mathbf{q}})+H.c.
	\bigg]  \nonumber\\
	&+\sum_{\rho\neq\rho^{\prime},\mathbf{q}}\sum_{m+l=\alpha+\beta}\frac{g_{\rho\rho^{\prime}}l}{2}\bigg[
	(-\phi^{\ast}_{\rho m} \phi^{\ast}_{\rho^\prime l})
	(\hat{\phi}_{\rho^\prime\alpha\mathbf{k}_1-\mathbf{q}} \hat{\phi}_{\rho\beta\mathbf{k}_1+\mathbf{q}}+\hat{\phi}_{\rho^\prime\alpha\mathbf{k}_1+\mathbf{q}} \hat{\phi}_{\rho\beta\mathbf{k}_1-\mathbf{q}})+(-\phi^{\ast}_{\rho m} \phi_{\rho^\prime -l})\times \nonumber\\
	&(\hat{\phi}^{\dagger}_{\rho^\prime -\alpha\mathbf{k}_1+\mathbf{q}}  \hat{\phi}_{\rho\beta\mathbf{k}_1+\mathbf{q}}+\hat{\phi}^{\dagger}_{\rho^\prime -\alpha\mathbf{k}_1-\mathbf{q}} \hat{\phi}_{\rho\beta\mathbf{k}_1-\mathbf{q}})
	+\phi^{\ast}_{\rho m} \phi_{\rho -l}
	(\hat{\phi}^{\dagger}_{\rho^\prime -\alpha\mathbf{k}_1+\mathbf{q}} \hat{\phi}_{\rho^\prime\beta\mathbf{k}_1+\mathbf{q}} +\hat{\phi}^{\dagger}_{\rho^\prime -\alpha\mathbf{k}_1-\mathbf{q}} \hat{\phi}_{\rho^\prime\beta\mathbf{k}_1-\mathbf{q}})+H.c.
	\bigg],
	\label{eq_H_B}
	\end{align}
\end{widetext}
where $m$, $l$, $i$, $j$, $\alpha$, $\beta$ are all odd integers with $-C_1 \leq m,n,i,j \leq C_1$ and $-C_2 \leq \alpha,\beta \leq C_2$, $C_2$ is another cutoff necessary for numerical diagonalization of Bogoliubov Hamiltonian, $q_x$ is in the first Brillioun zone, $0<q_x<k_1$, $n$ is total particle density, and $\mu=\sum\limits_{m}\sum\limits_{\rho,\rho^{\prime}}\bigg(\frac{k_r^2}{2}(m\gamma-\sigma_z)^2-\frac{\Omega}{2}\sigma_x\bigg)_{\rho\rho^{\prime}}\phi^{\ast}_{\rho m}\phi_{\rho^{\prime} m}
+\sum\limits_{m+l=i+j}\;\sum\limits_{\rho,\rho^{\prime}}g_{\rho\rho^{\prime}}n\phi^{\ast}_{\rho m}\phi^{\ast}_{\rho^\prime l}\phi_{\rho^{\prime} i}\phi_{\rho j}$ is, in nature, the MF chemical potential, satisfying
$\mu=\partial E_{\scriptscriptstyle M\!F}/\partial N$.

In Hamiltonian Eq.(\ref{eq_H_B}) there are not only terms equivalent to periodic potentials, but also off-diagonal terms such as $\hat{\phi}_{\rho \alpha\mathbf{k}_1+\mathbf{q}}\hat{\phi}_{\rho\alpha\mathbf{k}_1-\mathbf{q}}$. The quasiparticle spectra are characterized by band index and quasimomentum.	

\par We define a column operator
	\begin{align*}
	 \hat{A}_{\mathbf{q}}\equiv\big(\cdots,\hat{\phi}^{\dagger}_{\uparrow\alpha\mathbf{k}_1-\mathbf{q}},\hat{\phi}^{\dagger}_{\downarrow\alpha\mathbf{k}_1-\mathbf{q}},\hat{\phi}_{\uparrow\alpha\mathbf{k}_1+\mathbf{q}},\hat{\phi}_{\downarrow\alpha\mathbf{k}_1+\mathbf{q}},\cdots\big)^T
	\end{align*}
with $\alpha=\pm 1,\pm3,\dots,\pm C_2$,
and rewrite Eq.(\ref{eq_H_B}) in a compact form
\begin{align*}
H_{\scriptscriptstyle B}=E_{\scriptscriptstyle M\!F}+E_1+\sum_{\mathbf{q}}\hat{A}^{\dagger}_{\mathbf{q}}H_{\mathbf{q}}\hat{A}_{\mathbf{q}},
\end{align*}
where
\begin{align}
E_1&=-\sum_{m,\mathbf{q},\pm}\bigg[\frac{(m\mathbf{k}_1-\mathbf{q}\pm\mathbf{k}_r)^2}{2}-\mu+gn+\frac{g_{\uparrow\downarrow}n}{2}\bigg].
\label{eq_E_1}
\end{align}
\par The matrix $H_{\mathbf{q}}$ can be obtained from Eq.(\ref{eq_H_B}) and subsequently diagonalized to obtain quasiparticle spectra.  The diagonalized Bogoliubov Hamiltonian is given by
\begin{align}
&H_{\scriptscriptstyle B}=E_{\scriptscriptstyle M\!F}+E_1+\sum_{\alpha,\mathbf{q}}(E_{\alpha}^{\uparrow-}(\mathbf{q})+E_{\alpha}^{\downarrow-}(\mathbf{q}))\nonumber\\
&+\sum_{\alpha,\mathbf{q},\pm}\sum_{\rho}E_{\alpha}^{\rho\pm}(\mathbf{q})\hat{\tilde{\phi}}^{\dagger}_{\rho\alpha\mathbf{k}_1\pm\mathbf{q}}\hat{\tilde{\phi}}_{\rho\alpha\mathbf{k}_1\pm\mathbf{q}},
\label{eq_H_B_diagonal}
\end{align}
where $\big(\cdots,\hat{\tilde{\phi}}^{\dagger}_{\uparrow\alpha\mathbf{k}_1-\mathbf{q}},\hat{\tilde{\phi}}^{\dagger}_{\downarrow\alpha\mathbf{k}_1-\mathbf{q}},\hat{\tilde{\phi}}_{\uparrow\alpha\mathbf{k}_1+\mathbf{q}},\hat{\tilde{\phi}}_{\downarrow\alpha\mathbf{k}_1+\mathbf{q}},\cdots\big)^T=M_{\mathbf{q}}\hat{A}_{\mathbf{q}}$, $M_{\mathbf{q}}$ the Bogoliubov transformation matrix satisfying $M_{\mathbf{q}}\Sigma M_{\mathbf{q}}^{\dagger}=\Sigma$, and $\Sigma$ is a diagonal matrix with every four diagonal matrix elements given by $-1,-1,1,1$. The quasi-particle energy $E_{\alpha}^{\rho\pm}(\mathbf{q})$ can be solved from the generalized secular equation $\left|H_{\mathbf{q}}-\lambda\Sigma\right|=0$.

\par Before we write down the expression of LHY energy, we need to rewrite $g_{\rho\rho^\prime}$ in terms of scattering length $a_{\rho\rho^\prime}$ through regularization relation
$g_{\rho\rho^{\prime}}=U_{\rho\rho^{\prime}}+(U_{\rho\rho^{\prime}}^2/V)\sum\nolimits_{\mathbf{k}}1/k^2$ \cite{pitaevskii2016}
where $U_{\rho\rho^{\prime}}=4\pi a_{\rho\rho^{\prime}}$. The LHY energy is therefore given by
\begin{align}
&E_{\scriptscriptstyle L\!H\!Y}=E_1
+\sum_{\alpha,\mathbf{q}}\bigg[(E_{\alpha}^{\uparrow-}(\mathbf{q})+E_{\alpha}^{\downarrow-}(\mathbf{q}))+\bigg(\sum_{m+l=i+j}  \nonumber\\
&\sum_{\rho,\rho^{\prime}}\frac{(U_{\rho\rho^{\prime}}n)^{2}}{2}\phi^{\ast}_{\rho m}\phi^{\ast}_{\rho^\prime l}\phi_{\rho^\prime i}\phi_{\rho j}\bigg)\bigg(\sum_{\pm}\frac{1}{(\alpha\mathbf{k}_1\pm\mathbf{q})^2}\bigg)\bigg],
\label{eq_E_LHY}
\end{align}
where the summation converges quickly for large momentum due to regularization. In contrast, the divergence of LHY energy was removed by dimensional regularization in Ref. \cite{sanchez2020}.

\paragraph{Self-bound quantum droplet with stripe} With the explicit expressions of MF energy Eq.(\ref{eq_E_MF}) and LHY correction Eq.(\ref{eq_E_LHY}), we are ready to investigate the interplay between SOC, RC and interactions in the formation of droplet.  We take $E_r\equiv \mathbf{k}_r^2/2$ and $\mathbf{k}_r$ as energy and momentum units respectively.  The dimensionless version of MF and LHY energies are given in appendix.  In the following numerical calculations, we use the parameters $a_{\uparrow\uparrow}=a_{\downarrow\downarrow}\equiv a= 89.08a_0$, $a_{\uparrow\downarrow}=-1.1a$ where $a_0$ is the Bohr radius. Correspondingly, we define $U\equiv 4\pi a= U_{\uparrow\uparrow}=U_{\downarrow\downarrow}$. And before the effect of SOC is considered, the recoil momentum is fixed at $k_r=2\pi\times 10^6 m^{-1}$(or equivalently $ak_r\approx0.0296$).

\par When implementing the numerical calculations, we introduce two cutoffs: $C_1$ is for the ground-state ansatz, Eq.(\ref{eq_phi}), and $C_2$ is for the diagonalization of the otherwise infinite-dimensional Bogoliubov Hamiltonian, Eq.(\ref{eq_H_B}).  We have numerically verified the convergence of wavefunctions and LHY energies, and find that the choice of $C_1=9$ and $C_2=39$ can produce sufficiently accurate results.  The condensate fraction at $m=9$ is about $10^{-14}$ of the total density.  In our calculations, the two characteristic momenta, $\sqrt{gn}$ and $\sqrt{\Omega}$, are at most of the same order of the recoil momentum $k_r$.  The momentum cutoff $C_2 k_r$ is much larger than any of them.  Therefore, throughout our calculations, we set $C_1=9$ and $C_2=39$.

\begin{figure}
	\includegraphics[width=0.9\columnwidth]{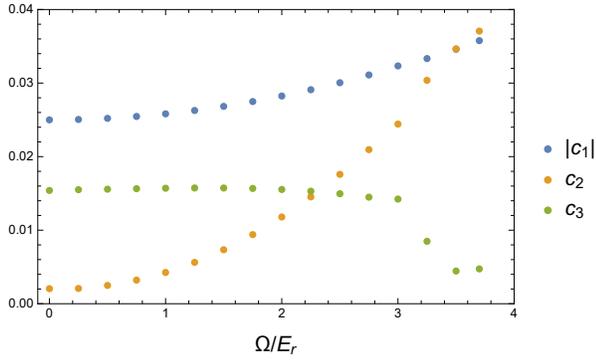}
	\caption{{\label{fig1}} Three fitting coefficients $c_1,c_2$ and $c_3$ in MF energy and LHY energy vs the strength of Raman coupling $\Omega$ at $ak_r\approx0.0296$. $c_1$ is negative and has been shown by its magnitude for convenient comparison with $c_2$.}
\end{figure}

We first minimize the mean-field energy at fixed total density $n$ to determine variational parameters $\phi_{\rho m}$ and $\gamma$.  It shows that, at low densities, the mean-field energy per particle $\varepsilon_{\scriptscriptstyle M\!F}$ shows linear dependence on density, and can be fitted by $\varepsilon_{\scriptscriptstyle M\!F}=c_0+c_1(Un/E_r)$. The density-independent background energy appears due to Raman energy in Eq.(\ref{eq_E_MF}), and thus $c_0$ depends strongly on the strength of RC, while the proportionality coefficient $c_1$, as shown in Fig.\ref{fig1}, weakly relies on the RC strength in the considered regime. Both $c_0$ and $c_1$ are irrelevant to the strength of SOC, $k_r$, since the MF energy has been rescaled in the unit of $E_r$. Moreover, $c_1$ is negative throughout the considered regime, which indicates the tendency to collapse in the mean field and thus higher-order correction is necessary to stabilize such a system.


\par In the low density region, the mean-field density distribution exhibits stripes as in the low RC limit in repulsive BEC-gas \cite{li2012}.   As has been studied \cite{lin2011,ji2014,li2012}, in experiments on $^{87}$Rb atoms, stripe phase exists only for very small $\Omega$ ($\lesssim 0.2E_r$), and stripes cannot be detected directly in the absoption imaging.   In contrast, the stripe phase  of the quantum droplet can survive in a much larger range of RC, of the order of several $E_r$.  This result can be obtained in 
the variational theory \cite{li2012} of a Bose gas with SOC, where the stripe phase and the plane-wave phase can all be described.  In the low density limit with strong SOC, there is a transition between these two phases at a critical RC, $\Omega^{\text{I-II}}=4E_r\sqrt{\frac{2\gamma}{1+2\gamma}}$ where $\gamma=(g-g_{\uparrow\downarrow})/(g+g_{\uparrow\downarrow})$ (see Eq.(12) in \cite{li2012}). Consequently, we can reach a conclusion that in a quantum droplet with $\Omega<4E_r$ and strong SOC, stripe phase is favored.  Compared to the BEC case with repulsive iner-species interaction where $\Omega^{\text{I-II}}\lesssim 0.2E_r$, one can draw the conclusion that it is the strong inter-species attraction that significantly enlarges the region of stripe phase.


\par We compute the LHY energy by solving excitation spectra and numerically performing the integration in Eq.(\ref{eq_E_LHY}).    As in the case without SOC \cite{petrov2015}, the lowest excitation spectrum at $q_x\approx 0$ and $2k_r$ has small imaginary part.    When performing the numerical integration, we keep all the real and imaginary contributions in the excitation energies.    Even though the resulting LHY energy is complex, the imaginary part is at least three orders smaller in magnitude than the real part in the parameter region that we are considering, i.e. with low density and strong SOC.   Consequently, the imaginary parts can be safely omitted in the ensuing calculations as in the case without SOC \cite{petrov2015}.

\par Notice that LHY energy is a function of three dimensionless parameters, $Un/E_r$, $\Omega/E_r$ and $ak_r$ (see Eq.(\ref{eq_ap_E_LHY})) whereas the MF energy depends only on the first two parameters as can be seen in Eq.(\ref{eq_ap_E_MF}). In the dilute limit, in terms of $Un/E_r$, the LHY energy per particle can be fitted by the formula $\varepsilon_{\scriptscriptstyle L\!H\!Y}=c_2(Un/E_r)+c_3(Un/E_r)^{3/2}$, where $c_2$ and $c_3$ are positive fitting parameters. Compared to the case without SOC, in addition to the usual term proportional to $n^{3/2}$, a linear repulsive term appears in the LHY energy which will be discussed later.

\par The coefficient $c_2$, as shown in Fig.\ref{fig1}, has a roughly quadratic dependence on RC strength while $c_3$ remains constant in the range of $0<\Omega<3E_r$, in agreement with the preceding work \cite{sanchez2020}.   Also, such a behavior occurs in the Rabi-coupled case \cite{cappellaro2017} where the coefficient of usual $n^{3/2}$ term in LHY energy is free of $\Omega$.   For $\Omega>3E_r$, only low density region can be sampled, and while the fitted $c_1$ and $c_2$ remains reliable, the value of $c_3$, which determines the behavior of the LHY energy in the higher density region, can not be trusted due to the deficiency of sampling.

\begin{figure}[hbt]
	\includegraphics[width=0.9\columnwidth]{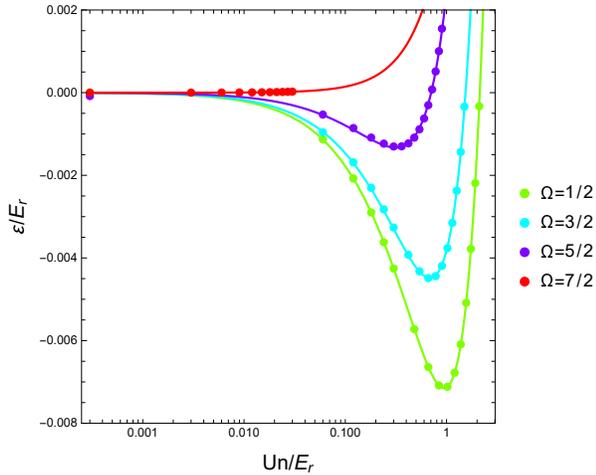}
	\caption{\label{fig2} Sampled points (dots) and fitting functions (solid lines) of total energy per particle $\varepsilon=\varepsilon_{\scriptscriptstyle M\!F}+\varepsilon_{\scriptscriptstyle L\!H\!Y}$ for several Raman coupling strengths at $ak_r\approx 0.0296$. The constant energy background from the MF energy has been subtracted for comparison.}
\end{figure}

\par Including both MF and LHY energies, the total energy per particle in the droplet regime is shown in Fig.\ref{fig2}. Near the collapse point of the MF energy, contrary to the monotonously decreasing tendency of the MF energy with density, the total energy has a minimum, because the repulsive LHY energy overcomes the attractive MF energy at larger densities. As discussed above, the MF wavefunction shows density modulation. Therefore, in the droplet regime, the self-bound stripe phase exists and can survive for even larger Raman coupling compared to the BEC gas regime \cite{lin2011,ji2014,li2012}.

\begin{figure}[hbt]
	\includegraphics[width=0.9\columnwidth]{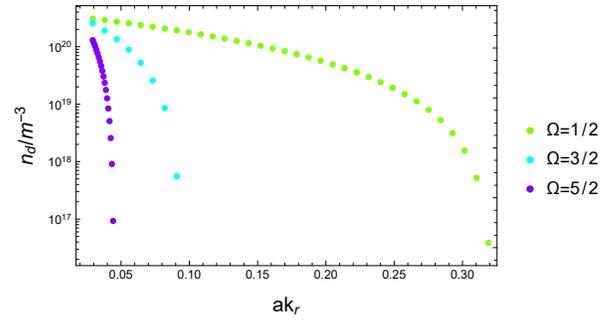}
	\caption{\label{fig3} Droplet density $n_d$ versus SOC strength $k_r$ for several RC strengths. With RC fixed, the stronger the SOC is, the smaller the equilibrium density of droplet becomes.  When SOC is strong enough, for example at $ak_r\approx 0.32$ for $\Omega=1/2E_r$, the droplet state disappears and the system expands without an external trap.}
\end{figure}

\par Although the general analytic total energy per particle is inaccessible in the presence of SOC and RC, it can be approximated by adding together the fitted MF and LHY energies, $\varepsilon=c_0+(c_2-\lvert c_1 \rvert)(Un/E_r)+c_3(Un/E_r)^{3/2}$ with $c_{0,1,2,3}$ all fitted numerically.  The equilibrium density of self-bound droplet can obtained by solving zero-pressure condition, i.e. $P=\partial\varepsilon/\partial V=0$, yielding $Un_{d}/E_r=4(\lvert c_1 \rvert-c_2)^2/9c_3^2$ for $\lvert c_1 \rvert-c_2\geq0$.

\par  We now discuss the role played by RC strength on droplet formation. The dependence of droplet on RC is similar to the case in uniform Rabi-coupled binary mixture which has been reported in 3D \cite{cappellaro2017} and lower dimensions \cite{chiquillo2019}. The similarity stems from the gapped single-particle spectrum. Without RC, the single-particle spectra of both components are gapless. The finite RC induces coupling between the two components leading to the new quasi-particle spectra, with the lower one gapless and the higher one gapped.  Due to the gapped mode, LHY correction per particle acquires a positive term linear in density $n$ in addition to the $n^{3/2}$ term, as mentioned above.

\par As $\Omega$ increases, the rapid increase of $c_2$ results in the decreasing of $\lvert c_1 \rvert-c_2$ in magnitude, which is manifest in Fig.\ref{fig1}, with $c_3$ remaining almost constant, making it easier to counterbalance the attractive MF energy.  Thus the equilibrium density of droplet is smaller for larger $\Omega$ as shown in Fig.\ref{fig2}.  A critical point is reached at $\lvert c_1 \rvert=c_2$, as shown in Fig.\ref{fig1} where $\Omega_c\approx3.5E_r$ at $ak_r\approx 0.0296$.  Above this point, the total energy increases monotonously with density since both $c_2-\lvert c_1 \rvert$ and $c_3$ are positive, and thereby no self-bound droplet can exist.  A similar droplet-gas transition has been reported theoretically in Rabi-coupled binary mixture in both 3D \cite{cappellaro2017} and lower dimensions \cite{chiquillo2019}.

\par It is easier to consider the dependence on SOC strength as $\mathbf{k}_r$ ($E_r\equiv \mathbf{k}_r^2/2$) serves as the momentum (energy) unit.   From the dimensionless expression of MF and LHY energies Eq.(\ref{eq_ap_E_MF}) and Eq.(\ref{eq_ap_E_LHY}), it's easy to see that the MF energy only depends on $Un/E_r$ and $\Omega/E_r$, and so does the excitation energy $E_{\alpha}^{\rho\pm}(\mathbf{q})$.   With fixed reduced interaction $Un/E_r$ and Raman coupling $\Omega/E_r$, the LHY energy is proportional to $k_r$, and as a consequence, the fitting parameters in LHY energy, $c_2$ and $c_3$ are both linearly proportional to $k_r$.   Since $c_1$ is independent of $k_r$, increasing SOC strength $k_r$ has the same effect as increasing RC strength $\Omega$.   Although the energy unit $E_r$ is also increased in the same process, the overall effect of increasing SOC strength is, as shown in Fig.\ref{fig3}, decreasing equilibrium density of droplet. And finally above some specific value, no droplet can exist any more.


\paragraph{Conclusion and Discussion}
\par In current experiment on $^{39}K$, the droplet state has been realized in the mixture of hyperfine states $\left|1,-1\right\rangle$ and $\left|1,0\right\rangle$ by tunning scattering lengths \cite{petrov2015,cabrera2018,cheiney2018,semeghini2018}, but the artificial SOC has not been realized in this system.  In contrast, in $^{87}Rb$ systems \cite{lin2011,zhang2012} the SOC has been realized and the stripe state has been observed, but tunning the interactions in this system has not been achieved. Our results could be tested experimentally if the interactions can be tuned and the artificial SOC can be generated in the same system.

In conclusion, we have studied the quantum droplet state of a uniform binary Bose gas in the presence of 1D spin-orbit-coupling and Raman coupling, and find that ground state can display density modulation of the stripe phase in the low $\Omega$ regime \cite{li2012,li2013}. The density modulation can survive for much larger $\Omega$ than in the BEC gas state with inter-species interaction. Compared to the case without SOC, the droplet density can be tuned by changing the strength of SOC and RC. With the increase of SOC and RC, the droplet density can be reduced by several orders of magnitude, and eventually to the zero limit at a critical $k_r$ or $\Omega$.  We plan to study the finite-size effect of the quantum droplet with SOC in the future work.

\appendix*
\subsection{Appendix}
\par Dimensionless MF and LHY energies per particle are given by
\begin{widetext}
	\begin{align}
	E_{\scriptscriptstyle M\!F}/(NE_r)=\sum_{m}\sum_{\rho\rho^{\prime}}\bigg((m\gamma-\sigma_z)^2-\frac{\Omega}{2}\sigma_x\bigg)_{\rho\rho^{\prime}}\phi^{\ast}_{\rho m}\phi_{\rho^{\prime} m}+\sum_{m+l=i+j}\;\sum_{\rho\rho^{\prime}}\frac{U_{\rho\rho^{\prime}}n}{2}\phi^{\ast}_{\rho m}\phi^{\ast}_{\rho^\prime l}\phi_{\rho^{\prime} i}\phi_{\rho j},
	\label{eq_ap_E_MF}
	\end{align}
\begin{align}
&E_{\scriptscriptstyle L\!H\!Y}/(NE_r)=(ak_r)\frac{1}{\pi^2(Un)}\sum_{\alpha}\int^{\gamma}_{0}\int^{\gamma C_2}_{-\gamma C_2}\int^{\gamma C_2}_{-\gamma C_2}dq_xdq_ydq_z
\bigg[(E_{\alpha}^{\uparrow-}(\mathbf{q})+E_{\alpha}^{\downarrow-}(\mathbf{q}))
-\sum_{\pm}\bigg((\alpha\gamma-q_x\pm 1)^2+q_y^2\nonumber\\
&+q_z^2-\mu+Un+\frac{U_{\uparrow\downarrow}n}{2}\bigg)
+\bigg(\sum_{m+l=i+j}\;\sum_{\rho\rho^{\prime}}\frac{U_{\rho\rho^{\prime}}n}{2}\phi^{\ast}_{\rho m}\phi^{\ast}_{\rho^\prime l}\phi_{\rho^{\prime} i}\phi_{\rho j}\bigg)\bigg(\frac{1/2}{(\alpha\gamma\pm  q_x)^2+q_y^2+q_z^2}\bigg)\bigg],
\label{eq_ap_E_LHY}
\end{align}
\end{widetext}
where $\Omega$, $U_{\rho\rho^\prime}n$ and excitation spectra $E^{\rho-}_{\alpha}$ are in the unit of $E_r$.


\bibliography{Quantum_Droplet}

\end{document}